\documentclass[a4paper,12pt]{article}
\usepackage{color}
\usepackage{amsfonts}
\usepackage{amsmath}
\usepackage{graphicx}
\usepackage{epsfig}
\usepackage{braket}
\usepackage{cancel}
\usepackage{pict2e}
\usepackage{axodraw4j}
\usepackage{pstricks}

\def\SU{\text{SU}}

\def\U{\text{U}}

\def\be{\begin{equation}}
\def\ee{\end{equation}}
\def\beq{\begin{equation}}
\def\eeq{\end{equation}}
\def\bea{\begin{eqnarray}}
\def\eea{\end{eqnarray}}

\def\<{\left\langle}
\def\>{\right\rangle}

\newcommand{\dfourz}{\mathrm{d}^{(4)}z}

\def\d{\mathrm{d}}
\allowdisplaybreaks[2]
\begin{document}
\bibliographystyle{OurBibTeX}
\begin{titlepage}
 \vspace*{-15mm}
\begin{flushright}
\end{flushright}
\vspace*{5mm}
\begin{center}
{ \sffamily \LARGE $A_4 \times \SU(5)$ SUSY GUT of Flavour in 8d}
\\[8mm]
T.~J.~Burrows\footnote{E-mail: \texttt{tjb54@soton.ac.uk}} and
S.~F.~King\footnote{E-mail: \texttt{king@soton.ac.uk}}
\\[3mm]
{\small\it
School of Physics and Astronomy,
University of Southampton,\\
Southampton, SO17 1BJ, U.K.
}\\[1mm]
\end{center}
\vspace*{0.75cm}
\begin{abstract}
\noindent We propose an $\SU(5)$ SUSY GUT-Family model with $A_4$ family symmetry in 8d
where the vacuum alignment is
achieved in an elegant way by the use of boundary conditions on orbifolds. 
The model involves $\SU(5)$ living in the 8d bulk,
with matter living in 6d (or 4d) subspaces and Yukawa interactions occurring at a 4d point.
The GUT group is broken to the Standard 
Model by the orbifold compactification, setting the GUT scale and 
leading to low energy supersymmetry and Higgs 
doublet-triplet splitting. The first two families of $\mathbf{10}$-plets are 
doubled resulting in a lack of both desirable and unwanted GUT relations. The resulting 
four dimensional effective superpotential leads to a realistic description of quark
and lepton masses and mixing angles including tri-bimaximal
neutrino mixing and an inter-family mass hierarchy
provided in part by volume suppression and in part by a Froggatt-Nielsen mechanism.
\end{abstract}
\end{titlepage}
\newpage
\setcounter{footnote}{0}

\section{Introduction}

It is well known that the solar and atmospheric data are
consistent with so-called tri-bimaximal (TB) mixing
\cite{HPS,Xing:2002sw},
\begin{eqnarray}
U_{TB} =
\left( \begin{array}{rrr}
-\frac{2}{\sqrt{6}}   & \frac{1}{\sqrt{3}} & 0 \\
\frac{1}{\sqrt{6}}  & \frac{1}{\sqrt{3}} & \frac{1}{\sqrt{2}} \\
\frac{1}{\sqrt{6}}  & \frac{1}{\sqrt{3}} & -\frac{1}{\sqrt{2}}
\end{array}
\right).
\label{MNS0}
\end{eqnarray}
The ansatz of TB lepton mixing matrix is interesting due to its
symmetry properties which seem to call for a possibly discrete
non-Abelian Family Symmetry in nature \cite{Harrison:2003aw}.
There has been a considerable amount of theoretical work in which
the observed TB neutrino flavour symmetry may be related to some
Family Symmetry
\cite{Chen:2009um,King:2005bj,deMedeirosVarzielas:2005ax,deMedeirosVarzielas:2006fc,King:2006me,
Altarelli:2006ri,Frampton:2004ud,King:2009mk,Lam:2008sh,Grimus:2009pg}.
These models may be classified according to the way that TB mixing
is achieved, namely either directly or indirectly
\cite{King:2009ap}. The direct models are based on $A_4$ or $S_4$,
or a larger group that contains these groups as a subgroup, and in
these models some of the generators of the Family Symmetry survive
to form at least part of the neutrino flavour symmetry. In the
indirect models, typically based on $\Delta(3n^2)$ or
$\Delta(6n^2)$, none of the generators of the Family Symmetry
appear in the neutrino flavour symmetry \cite{King:2009ap}.
All of the above models rely on some kind of vacuum alignment mechanism,
which in 4d SUSY models may arise from either F-terms in direct models or D-terms 
in indirect models (for a recent discussion see e.g. \cite{Howl:2009ds}).

The most ambitious models combine Family Symmetry with Grand
Unified Theories (GUTs). The minimal Family Symmetry which
contains triplet representations and can lead to TB mixing via the
direct model approach is $A_4$. The minimal simple GUT group is
$SU(5)$. A direct model has been proposed which combines $A_4$
Family Symmetry with $SU(5)$ Supersymmetric (SUSY) GUTs
\cite{Altarelli:2008bg}. This model was formulated in five
dimensions (5d), in part to address the doublet-triplet splitting
problem of GUTs \cite{Asaka:2001eh}, and in part to allow a viable description of the
charged fermion mass hierarchies, by placing the lightest two
ten-plets $T_1$, $T_2$ in the bulk, while the pentaplets $F$ and
$T_3$ are on the brane. It was subsequently shown how the geometry of
6d compactification may be used to generate the $A_4$ symmetry dynamically \cite{Altarelli:2006kg}
(see also \cite{Adulpravitchai:2009id,Kobayashi:2006wq})
and subsequently an $SU(5)$ SUSY GUT model in 6d was proposed in which the 
$A_4$ symmetry arises dynamically \cite{Burrows:2009pi}. However, 
in both the 5d model \cite{Altarelli:2008bg} and the 6d model \cite{Burrows:2009pi},
the $A_4$ was broken at the effective 4d level in the standard way using F-terms to align
the two flavons $\varphi_T$, $\varphi_S$ via the 
introduction of so-called driving fields.

In the framework of extra dimensional theories, an attractive
alternative mechanism for vacuum alignment arises based on orbifolding
\cite{Kobayashi:2008ih}. The required alignment at the zero mode level 
is achieved by imposing non-trivial boundary conditions on the orbifold \cite{Kobayashi:2008ih}. 
In order to achieve the desired vacuum alignment for 
the two flavons $\varphi_T$, $\varphi_S$ in an $A_4$ model it was 
demonstrated that it is necessary to formulate the model in 
8 dimensions, which allows 4 compact dimensions which may be regarded as  
two complex compact dimensions $z_1, z_2$, where 
each is subject to a particular 2d orbifold which gives the
vacuum alignment for the particular flavon $\varphi_T$, $\varphi_S$
living in those dimensions \cite{Kobayashi:2008ih}. 
This mechanism has also been applied to $S_4$ where it has been shown that 
some of the desired alignments may be achieved in 6d \cite{Adulpravitchai:2010na}.
However, in the case of  $A_4$ it is clear that it is necessary to formulate the
model in 8d in order to achieve successful vacuum alignments via boundary conditions,
although so far only an illustrative model along these lines
has been presented \cite{Kobayashi:2008ih}.

The purpose of this paper is to formulate the first realistic 
$\SU(5)$ SUSY GUT model with $A_4$ family symmetry in 8d
where the vacuum alignment is straightforwardly 
achieved by the use of boundary conditions on orbifolds of the four compact
dimensions. We emphasise that we are motivated to consider an 8d theory by the desire to
achieve vacuum alignment in an elegant way using orbifold boundary conditions.
It is not possible to implement this idea with lower dimensional models such as the
the 5d model in \cite{Altarelli:2008bg} or the 6d model in \cite{Burrows:2009pi}
since the desired alignment mechanism is not possible 
under a single orbifolding due to the requirement that 
the two triplet flavons $\varphi_T$ and $\varphi_S$ have different boundary 
conditions in order to have the different alignments at the zero mode level.  
Working in 8d also brings
additional benefits, for example the inter-family mass hierarchies will arise in part due to suppression 
factors arising from an asymmetric geometric dilution of the wavefunctions in the four compact
dimensions, although a $U(1)$ Froggatt-Nielsen family symmetry will also be required. 
In the 8d model the 4 extra dimensions are compactified onto 2 complex 
directions which are each orbifolded with $\mathbb{Z}_2$ and $\mathbb{Z}_3$ 
symmetries. These orbifoldings are also used to specify non-trivial 
boundary conditions on the various multiplets which break the $\SU(5)$ gauge symmetry and 
the extended $\mathcal{N}=4$ symmetry to leave an effective $\mathcal{N}=1$ Standard Model 
theory in 4 dimensions. It is worth noting that due to the orbifoldings 
the first two families of $\mathbf{10}$-plets are duplicated introducing new GUT scale 
mass particles to the theory, although such a feature removes any 
desirable GUT predictions it also removes some unwanted GUT mass relations. 


The layout of the remainder of the paper is as follows. In Section \ref{sec:model} we 
introduce the model and show how the 8 dimensions are compactified upon 
two $\mathbb{T}^2/(\mathbb{Z}_2\times\mathbb{Z}_3)$ orbifolds leading to gauge and SUSY breaking
as above. We specify the 
superfield content and symmetries of the model.  
We describe the transformation of the fields under these 
orbifoldings which leads to an effective 4d Standard Model theory from the 
8d $\SU(5)$ theory. We first show how the GUT group is broken and 
how this naturally leads to doublet-triplet splitting of the Higgs multiplets.
We then discuss vacuum alignment in the 8d theory, and show how boundary conditions can lead
to the desired alignment directions.
We also discuss the values of the Higgs and flavon VEVs, including the effects of
bulk suppression factors. In Section \ref{sec:superpotentials} we write down the effective 4d superpotential 
and the resulting mass matrices. We also analyse contributions from terms beyond the 
leading order to the mass matrices. Section \ref{sec:conclusion} concludes the paper.

\section{The Model}
\label{sec:model}

We are considering a model in 8 dimensions with the extra dimensions compactified on two 2d orbifolds 
as described in sec. \ref{sec:orbifold}. 
The $\SU(5)$ gauge group lives in the full 8d bulk, with 
the 8d space compactified to 4d Minkowski space $\times$ 4d compact dimensions with the 
two complex compact dimensions described by the coordinates $z_1$ and $z_2$.
We suppose that the 8d space is compactified by orbifolding.
In the $z_1$ direction the $\mathbb{Z}_2$ orbifolding breaks the gauge 
symmetry and gives the alignment of the $A_4$ flavon $\varphi_S$, 
while the $\mathbb{Z}_3$ orbifold breaks the extended 
supersymmetry as described below. 
In the $z_2$ direction the $\mathbb{Z}_2$ orbifolding also breaks the gauge symmetry 
to the Standard Model in exactly the same way as in the $z_1$ direction, 
while the $\mathbb{Z}_3$ symmetry is used to 
give the alignment of the $A_4$ flavon $\varphi_T$ as described in sec. \ref{sec:alignment} 
and \cite{Kobayashi:2008ih}.

We suppose that some of the matter and Higgs fields 
do not feel the full 8d but are restricted to live in a 6d subspace
of the full 8d theory. The second family of $\mathbf{10}$'s, $T_2$,
live in the $z_1$ direction along with both Higgs multiplets, $H_5$ and $H_{\overline{5}}$. 
The first family of $\mathbf{10}$'s, $T_1$, is placed in the $z_2$ direction. 
Similarly, the flavons $\varphi_S$, $\xi$ and $\theta''$ live in the  $z_1$ direction,
with $\varphi_T$ and $\theta$ in the  $z_2$ direction. 
We confine the other
matter fields to live in a 4d subspace, with the three families of 
$\bar{\mathbf{5}}$ matter, $F$, and the third family of $\mathbf{10}$'s, $T_3$, 
along with the three families of right-handed neutrinos, $N$,
located at the 4 dimensional fixed point $z_1=z_2=0$, with 
the Yukawa couplings given by the overlap of the wavefunctions at this fixed point. 
The particle content of the model is summarised in table \ref{table:particles}.

\begin{figure}[h]
\begin{center}
\scalebox{0.6}{
\fcolorbox{white}{white}{
  \begin{picture}(544,384) (211,-107)
    \SetWidth{1.0}
    \SetColor{Black}
    \Line[arrow,arrowpos=1,arrowlength=10,arrowwidth=4,arrowinset=0.2](352,-80)(352,224)
    \Line[arrow,arrowpos=1,arrowlength=10,arrowwidth=4,arrowinset=0.2](288,-32)(688,-32)
    \Vertex(352,-32){2.828}
    \Text(336,256)[lb]{\Large{\Black{$\mathbb{T}^2/(\mathbb{Z}_2^{\mathrm{SM}}\times\mathbb{Z}_3)$}}}
    \Text(720,-32)[lb]{\Large{\Black{$\mathbb{T}^2/(\mathbb{Z}_2^{\mathrm{SM}}\times\mathbb{Z}^{\mathrm{SUSY}}_3)$}}}
    \Text(608,-64)[lb]{\Large{\Black{$T_2,{H}_{\bar{\mathbf{5}}},H_{\mathbf{5}}$}}}
    \Text(560,-16)[lb]{\Large{\Black{$\varphi_S,\xi,\theta''$}}}
    \Text(368,128)[lb]{\Large{\Black{$\varphi_T,T_1,\theta$}}}
    \Text(330,200)[lb]{\Large{\Black{$z_2$}}}
    \Text(670,-20)[lb]{\Large{\Black{$z_1$}}}
    \Text(250,-80)[lb]{\Large{\Black{$F,T_3,N$}}}
    \Line[arrow,arrowpos=1,arrowlength=10,arrowwidth=4,arrowinset=0.2](300,-60)(343,-38)  
   \Text(560,128)[lb]{\Huge{\Black{$\SU(5)$}}}
\end{picture}
}}
\end{center}
\caption{\small A Schematic diagram of the model. The $SU(5)$ gauge group is in the 8d bulk,
represented here by the entire $(z_1,z_2)$ plane,
while matter and Higgs fields are confined to 6d subspaces, represented by the complex coordinate directions 
$z_1$ and $z_2$, or to the 4d subspace, represented by the point at the origin.
The First and Second families are placed in the 
$z_2$ and $z_1$ directions respectively. Because there is a gauge breaking orbifolding, in both directions, 
half of the $\mathbf{10}$-plets become heavy so aditional multiplets are introduced in both directions with opposite 
parity to obtain the full SM particle content.}
\label{fig:model}
\end{figure}
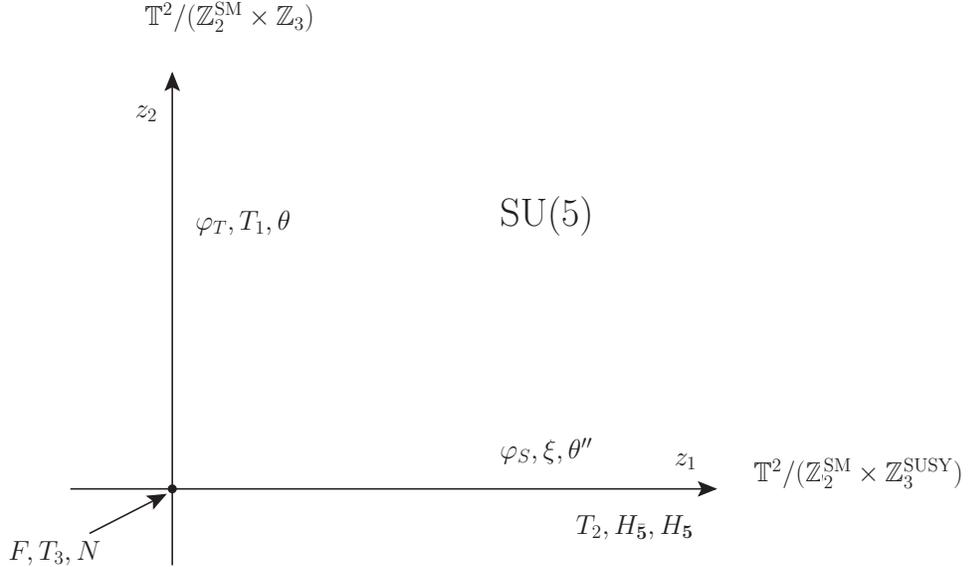

\begin{center}
\begin{table}[h]
\scalebox{0.85}{
 \begin{tabular}{|c|c|c|c|c|c|c|c|c|c|c|c|c|}
   \hline
   Superfield & N & F & $T_1$ & $T_2$ & $T_3$ & $H_5$ & $H_{\overline{5}}$ & $\varphi_T$&$\varphi_S$ &$\xi$&
   $\theta$&$\theta''$ \\
    \hline
    SU(5) & 1 & $\overline{5}$ & 10 & 10 & 10 & 5 & $\overline{5}$& 1 & 1 & 1 & 1 & 1 \\
    SM & 1 & $\scriptstyle(d^c,l)$& $\scriptstyle({u}^c_1,{q}_1,{e}^c_1)$& $\scriptstyle({u}^c_2,{q}_2,{e}^c_2)$&$\scriptstyle (u^c_3,q_3,e^c_3)$& $H_u$ &$H_d$ &$\varphi_T$ &$\varphi_S$ & $\xi$& $\theta$&$\theta''$\\
    $A_4$ & 3 & 3 & $1''$ & $1'$ & 1 & 1 & $1'$ & 3 & 3 & 1 & 1 & $1''$ \\
    $U(1)$ & 0 & 0 & 2 & 1 & 0 & 0 & 0 & 0 & 0 & 0 & -1 & -1 \\
    $\mathbb{Z}_3$ & $\omega$ & $\omega$ & $\omega$& $\omega$& $\omega$& $\omega$& $\omega$ & 1 & $\omega$& $\omega$ & 1 & 1 \\
   $U(1)_R$ & 1 & 1& 1& 1& 1& 0 & 0 & 0 & 0 & 0 & 0 & 0\\
   \hline
  Location & \tiny $z_1=z_2=0$ &\tiny $z_1=z_2=0$ &\tiny $z_1=0$&\tiny $z_2=0$ &\tiny $z_1=z_2=0$&\tiny $z_2=0$ &\tiny $z_2=0$ &\tiny $z_1=0$ &\tiny $z_2=0$ &\tiny $z_2=0$ &\tiny $z_1=0$ &\tiny $z_2=0$  \\
  \hline
   \end{tabular}
}
\caption{\small The Particle content and symmetries of the model.} 
\label{table:particles}
\end{table}
\end{center}

A schematic diagram of the model is shown in figure \ref{fig:model}.
As both the $z_1$ and $z_2$ directions have a $\mathbb{Z}_2$ orbifolding breaking the gauge symmetry, doublet-triplet splitting of the Higgs multiplets occurs. However this results in 
half the $\mathbf{10}$-plet becoming heavy. To overcome this, 
an extra copy of $\mathbf{10}$'s must be included 
in both directions with opposite parity under the $\mathbb{Z}_2$ symmetry. This 
results in the complete matter content and also allows us to escape unwanted GUT mass relations. 
In addition to the unwanted GUT mass relations the doubling of the first two families 
also prevents good GUT predictions such as the Gatto-Sartori-Tonin and Georgi-Jarlskog relations.
The 8 dimensional theory has an $A_4$ family symmetry which is broken by three 
flavons $\varphi_T,\varphi_S$ and $\xi$. 
The vacuum alignment of the flavons is achieved by imposing  non-trivial 
boundary conditions on the flavons so that only the required alignment 
has a zero-mode. In addition to the $A_4$ flavour symmetry there is volume suppression
for superpotential terms
involving 6d fields.
This suppression, however, turns out to be insufficient to account for realistic masses and 
mixings. To obtain a realistic pattern we also exploit the 
Froggatt-Nielsen mechanism \cite{Froggatt:1978nt}
with a $U(1)$ symmetry and 
the two Froggatt-Nielsen flavons $\theta$ and $\theta''$
living in the different orbifolded directions.
We also make use of $\U(1)_R$ and $\mathbb{Z}_3$ symmetries as shown in table \ref{table:particles}.

\subsection{The $\mathbb{T}^2/(\mathbb{Z}_2\times\mathbb{Z}_3)$ orbifolds}
\label{sec:orbifold}
The orbifolding can be 
used to break both the gauge symmetry and SUSY \cite{Asaka:2001eh}.
Models have also been proposed \cite{Burrows:2009pi} that combine these 
two ideas to give an extra dimensional GUT theory with a family symmetry arising from the compactification of the extra dimensions.
In the present model we will not insist that the family symmetry is dynamically generated from the 
compactified geometry of extra dimensions, but merely suppose that it pre-exists in the 8d theory.
However the part of the 
orbifold $\mathbb{T}^2/\mathbb{Z}_2$ described in this section is the same as that described in 
\cite{Altarelli:2006kg,Burrows:2009pi} where the $A_4$ is dynamically generated.  
The new feature here is that we shall use orbifold boundary conditions to give the desired vacuum alignment
for the flavons which break $A_4$, thereby yielding TB neutrino mixing.
We complexify the extra dimensions $x_5,x_6$ so that they are described by 
one complex coordinate $z_1=x_5+ix_6$. The extra dimensions are compactified on 
the a twisted torus defined by identifying the following translations:
\begin{eqnarray}
\label{eqn:torus1}
 z_1 \rightarrow z_1+1\\
\label{eqn:torus2}
 z_1 \rightarrow z_1+\gamma
\end{eqnarray}
where $\gamma=e^{i\pi/3}$ and we have set $2\pi R_{z_1}$, the length of the extra dimension, to unity. 
We then impose the following identification:
\begin{equation}
\label{eqn:z2}
 \mathbb{Z}_2: z_1 \rightarrow -z_1.
\end{equation}
 This defines the orbifold $\mathbb{T}^2/\mathbb{Z}_2$ as in \cite{Altarelli:2006kg,Burrows:2009pi}. We can also impose a $\mathbb{Z}_3$ symmetry 
in order to define the orbifold $\mathbb{T}^2/(\mathbb{Z}_2\times\mathbb{Z}_3)$, we impose the following identification:
\begin{equation}
\label{eqn:z3}
\mathbb{Z}_3 : z_1 \rightarrow \omega z_1.
\end{equation} 
Combining eqns. (\ref{eqn:torus1})-(\ref{eqn:z3}) 
gives the definition of the orbifold $\mathbb{T}^2/(\mathbb{Z}_2\times\mathbb{Z}_3)$ 
which is the complex direction denoted by $z_1$ in figure \ref{fig:model}. We follow an analogous
procedure for the remaining 2 
extra dimensions by defining $z_2=x_7+ix_8$ and imposing the above definitions substituting $z_2$ for $z_1$.
In other words, we apply $\mathbb{T}^2/(\mathbb{Z}_2\times\mathbb{Z}_3)$ orbifolding separately
in each of the $z_1$ and $z_2$ spaces.
The overall orbifold has a single fixed point invariant under both the $\mathbb{Z}_2$ and $\mathbb{Z}_3$ transformations 
which is located at $z_{1,2}=0$. It is at this 4d point that the Yukawa interactions occur.

\subsection{SUSY Breaking}
The full 8d theory is $\mathcal{N}=1$ $\SU(5)$ and the 8d bulk of the theory contains the $\SU(5)$ 
gauge bosons. Because spinors in 8 dimensions contain a minimum of 
16 real components then in 4 dimensions the effective theory must 
have $\mathcal{N}=4$ supersymmetry \cite{ArkaniHamed:2001tb}. 
In order to eliminate this extended supersymmetry 
we can impose boundary conditions on the multiplets so that they become heavy and play no part in the 
zero mode physics. The $\mathcal{N}=4$ vector multiplet 
decomposes into 3 chiral $\phi_i$ and one vector $V$ $\mathcal{N}=1$ multiplets. 
We can use the $\mathbb{T}^2/\mathbb{Z}_3$ 
part of the orbifolding to eliminate the 
unwanted multiplets by imposing the boundary 
conditions:
\begin{eqnarray}
 V(x^\mu,z_1,z_2)&=& V(x^\mu,\omega z_1,z_2)\\
 \phi_i(x^\mu,z_1,z_2)&=&\omega \phi_i(x^\mu,\omega z_1,z_2),
\end{eqnarray}
where $\omega$ are the cube roots of unity,
 leaving $\phi=0$ at the fixed point at $z_{1,2}=0$. We are therefore left with an effective $\mathcal{N}=1$ theory 
in 4 dimensions. 

\subsection{Gauge breaking through orbifolding}
 
The breaking of the $SU(5)$ gauge group
down to that of the Standard Model can be
achieved by the $\mathbb{Z}_2$ part of the orbifolding. 
By using a single parity $P_{SM}$,
\begin{equation}
 P_{SM}=\begin{pmatrix}
         +1 & 0 & 0 & 0 & 0 \\
          0 & +1 & 0 & 0 & 0 \\
          0 & 0 & -1 & 0 & 0 \\
          0 & 0 & 0 & -1 & 0 \\
          0 & 0 & 0 & 0 & -1 \\
        \end{pmatrix}
\end{equation}
we shall require that:
\begin{equation}
 P_{SM}V_\mu(x,-z)P^{-1}_{SM}=+V_\mu(x,z).
\end{equation}
Gauge boson fields of the standard model thus have positive parity
and fields belonging to $SU(5)/G_{SM}$ have negative parity.
Only fields with a positive parity have zero modes and therefore
gauge bosons not belonging to the standard model gauge group become heavy
and the gauge symmetry is broken. In our model both the $z_1$ and $z_2$ 
directions are orbifolded in this way, this allows us to relax unwanted GUT 
relations between the down quark and charged lepton mass matrices.

\subsection{Higgs and doublet-triplet splitting}

So far we have just considered the gauge sector of $SU(5)$.
Adding the Higgs to the 6d theory is straightforward.
In the $SU(5)$ GUT theory these are contained in
the $\textbf{5}$-plet and $\overline{\textbf{5}}$-plet of Higgs
fields. For the gauge breaking orbifold we choose:
\begin{eqnarray*}
 P_{SM}H_{\mathbf{5}}(x,-z_1)&=&+H_{\mathbf{5}}(x,z_1)\\
P_{SM}H_{\bar{\mathbf{5}}}(x,-z_1)&=&+H_{\bar{\mathbf{5}}}(x,z_1)
\end{eqnarray*}
It is easy to see with the form of $P_{SM}$ that the last three
entries gain a minus sign which makes them heavy whereas the first
two entries are left unchanged leaving them light, resulting in a
light doublet and a heavy coloured triplet. Similarly with the $\mathbf{10}$-plets 
living in the $z_1$ and $z_2$ directions half the multiplet becomes 
heavy, however by introducing extra multiplets with opposite 
parity the full particle content is restored at zero mode. This feature 
also allows us to evade unwanted GUT relations.

\subsection{Vacuum Alignment, VEVs and Expansion Parameters}
\label{sec:alignment}
\label{sec:vev}

In order to break the $A_4$ family symmetry we will impose non-trivial boundary conditions 
on flavons under the orbifoldings so that only a particular alignment survives to low energy. By imposing 
boundary  conditions we are able to avoid introducing the driving fields and avoid having to
write down a possibly complicated flavon potential. We will now describe the procedure for obtaining the 
alignment, closely following the procedure developed in \cite{Kobayashi:2008ih} to which we refer the
reader for more details. The first $\mathbb{Z}_2$ boundary condition,
\begin{equation}
 \varphi_S(-z_1)=P_2\varphi_S(z_1),
\end{equation}
requires the matrix $P_2$ to be of order 2. For $A_4$ we have the elements in the 
fourth conjugacy class to choose from. We can choose the matrix $P_2=S$ where $S$
is given by
\begin{equation}
 S=\begin{pmatrix}
    1 & 0 & 0 \\
    0 & -1 & 0 \\
    0 & 0 & -1
   \end{pmatrix}
\end{equation}
in the basis of $A_4$ where S is diagonal. This makes it trivial to 
see which alignment is left as a zero mode. This choice leaves a single zero mode in the $(1,0,0)$ direction in 
this basis. To find what this alignment is in the T diagonal basis it is a simple matter to
rotate the vector using  (for example see \cite{Burrows:2009pi}):
\begin{equation}
 V=\frac{1}{\sqrt{3}}\begin{pmatrix}
                       1 & \omega & \omega^2\\
                       1 & \omega^2 & \omega\\
                       1 & 1 & 1 
                      \end{pmatrix},
\end{equation}
This leaves us with the alignment $\varphi_S\propto(1,1,1)$ in the T diagonal basis.
For the $\mathbb{Z}_3$ orbifolding we can impose the boundary condition
\begin{equation}
 \varphi_T(\omega z_2)= P_3\varphi_T(z_2)
\end{equation}
and we can choose $A_4$ elements which have order 3. For $P_3$ we choose $P_3=T$ 
where 
\begin{equation}
  T =\begin{pmatrix}
           1 & 0 & 0 \\
           0 & \omega & 0 \\
           0 & 0 & \omega^2
      \end{pmatrix},
\end{equation}
This gives a single zero mode $\varphi_T\propto(1,0,0)$.

Turning to the VEVs themselves, for simplicity from now on
we shall set the radii of the compact directions to $R_5=R_6=R_{z_1}$ and $R_7=R_8=R_{z_2}$,
which implies that the Higgs VEVs are given by

\begin{equation}
 \braket{H_u(z_2)}=\frac{v_u}{\sqrt{\pi^2 R_{z_1}^2\sin\theta}},
 \ \ \braket{H_d(z_1)}=\frac{v_d}{\sqrt{\pi^2 R_{z_1}^2\sin\theta}}
\end{equation}
where we have included the effect of arbitrary 
twist angle $\theta$ on the torus \cite{Burrows:2009pi}.
For numerical estimates we will set the twist angle to $60^\circ$ 
(by choosing $\gamma=e^{i\pi/3}$ in eqn. \ref{eqn:torus2}) as in 
 \cite{Burrows:2009pi} (although in the present model this is an arbitrary choice).  
 
A useful feature of this setup is the suppression of the Yukawa couplings of fields living in the bulk. A 
field living in the 6d bulk of one of the orbifolded directions is related to it's zero mode by
\begin{equation}
 F(x^\mu,z)=\frac{1}{\sqrt{V}}F^0+\dots 
\end{equation}
where the dots represent the higher, heavy modes and V is the volume of the extra dimensional space. The 
above expansion produces a factor $s$:
\begin{equation}
 s=\frac{1}{\sqrt{V\Lambda^2}}.
\end{equation}
This feature will produce suppression for couplings 
involving these bulk fields. Since we are considering 6 dimensional fields that live in either the 
$z_1$ or $z_2$ direction we will have two not necessarily equal volume factors, $s_1$ and $s_2$:

\begin{equation}
 s_1=\frac{1}{\sqrt{\pi^2R_{z_1}^2\sin\theta\Lambda^2}}=\frac{1}{\sqrt{V_{z_1}\Lambda^2}}< 1
\label{eqn:suppression1}
\end{equation}
and 
\begin{equation}
 s_2=\frac{1}{\sqrt{\pi^2R_{z_2}^2\sin\theta\Lambda^2}}=\frac{1}{\sqrt{V_{z_2}\Lambda^2}}< 1.
\label{eqn:suppression2}
\end{equation}

Including volume suppression factors, we summarise the aligned flavon VEVs as follows,

\begin{eqnarray}
\frac{\braket{\varphi_{T}}}{\Lambda}&=&\frac{1}{\sqrt{V_{z_2}}}(v_T,0,0), \label{phi}\\ 
\frac{\braket{\varphi_S}}{\Lambda}&=&\frac{1}{\sqrt{V_{z_1}}}(v_S,v_S,v_S),\\
 \frac{\braket{\xi}}{\Lambda}&=&\frac{1}{V_{z_1}}u, \\
 \frac{\braket{\theta}}{\Lambda}&=&\frac{1}{\sqrt{V_{z_2}}}t,\\
  \frac{\braket{\theta''}}{\Lambda}&=&\frac{1}{\sqrt{V_{z_1}}}t'' 
  \label{theta}
\end{eqnarray}
We have defined the parameters $v_T,v_S,t$ and $t''$ so that they are dimensionless recalling that 6d fields have mass dimension two. The Froggatt-Nielsen flavons $\theta,\theta''$ require no special vacuum alignment and 
are assumed to obtain VEVs $t,t''$ of 
$\mathcal{O}(1)$. Such 
VEVs can be obtained as in \cite{Altarelli:2008bg} by minimising the D-term scalar potential. Obtaining VEVs of 
$\mathcal{O}(1)$ can be found by assuming appropriate mass and coupling parameters.

\section{Superpotentials and Mass Matrices}
\label{sec:superpotentials}
The couplings are localised at the single fixed point located at $z_1=z_2=0$ in the 
extra dimensional space. The action reads 

\begin{equation}
\int \d^4x\int \dfourz \int \d^2\theta w(x)\delta(z_1)\delta(z_2)+ h.c.= \int \d^4x \int \d^2\theta w(x)+h.c.
\end{equation}
The effective superpotential $w$ is expressed in terms of $\mathcal{N}=1$ superfields can be decomposed into the 
following parts:
\begin{equation}
 w=w_{\mathrm{up}}+w_{\mathrm{down}}+w_{\mathrm{charged\ lepton}}+w_\nu+w_{\mathrm{flavon}}
\end{equation}
The fermion masses and mixings are given by the first three parts after $A_4$,$U(1)$ Froggatt-Nielsen and electroweak symmetry 
breaking. The $w_{\mathrm{flavon}}$ part concerns the flavon fields, however since the $A_4$ flavon alignment is given by 
the non-trivial boundary conditions imposed by the orbifolding we can avoid writing down explicitly the (possibly complicated) 
flavon potential. However without explicitly writing the flavon potential we do lose the ability 
to make specific claims on relations between the $A_4$ flavon VEVs.

\subsection{Superpotentials}
We shall now write down the superpotentials of the model (excluding $w_\nu$ which is discussed in sec. \ref{sec:neutrino}). We shall use Standard Model notation since the theory 
is broken to the Standard Model gauge group by the compactification. We have suppressed the coefficients 
in each term of the superpotentials and we would expect such coefficients to be of $\mathcal{O}(1)$. We shall use the notation 
for fields $(f)'$ where the field transforms as a $\mathbf{1}'$ and similarly $(f)''$ for a $\mathbf{1}''$ of $A_4$.

\begin{eqnarray}
 w_{\mathrm{up}}&\sim&\frac{1}{\Lambda}H_uq_3u^c_3+\frac{{\theta''}}{\Lambda^{4}}H_u\{(q_2)'u^c_3+q_3({u}^c_2)'\}+\frac{{\theta''}^2}{\Lambda^{7}}H_u\{(q_2)'({u}^c_2)'\}\nonumber\\
&+& \frac{{\theta''}^2}{\Lambda^{6}}H_u\{(q_1)''u^c_3+q_3({u}^c_1)''\}+ \frac{{\theta''}^3+\theta^3}{\Lambda^{9}}H_u\{(q_2)'({u}^c_1)''+(q_1)''({u}^c_2)'\} \nonumber \\
&+&\frac{{\theta''}\theta^3+{\theta''}^4}{\Lambda^{11}}H_u\{(q_1)''({u}^c_1)''\},
\label{eqn:model2.1}
\end{eqnarray}

\begin{eqnarray}
 w_{\mathrm{down}}&\sim&\frac{1}{\Lambda^{3}}(H_d)'(d^c\varphi_T)''q_3+\frac{{\theta''}}{\Lambda^{6}}(H_d)'(d^c\varphi_T)''(q_2)' + \frac{{\theta}}{\Lambda^{6}}(H_d)'(d^c\varphi_T)'(q_2)' \nonumber \\
         &+& \frac{{\theta''}^2}{\Lambda^{8}}(H_d)'(d^c\varphi_T)''(q_1)''\nonumber \\
         &+& \frac{{\theta''}\theta}{\Lambda^{8}}(H_d)'(d^c\varphi_T)'(q_1)''+\frac{\theta^2}{\Lambda^{8}}(H_d)'(d^c\varphi_T)(q_1)''
\label{eqn:model2.2}
\end{eqnarray}

\begin{eqnarray}
 w_{\mathrm{charged\ lepton}}&\sim& \frac{1}{\Lambda^{3}}(H_d)'(l\varphi_T)''e^c_3 +\frac{{\theta''}}{\Lambda^{6}}(H_d)'(l\varphi_T)''({e^c}_2)'+\frac{{\theta}}{\Lambda^{6}}(H_d)'(l\varphi_T)'({e^c}_2)' \nonumber \\
 &+& \frac{{\theta''}^2}{\Lambda^{8}}(H_d)'(l\varphi_T)''({e^c}_1)'' \nonumber \\
 &+& \frac{{\theta''}\theta}{\Lambda^{8}}(H_d)'(l\varphi_T)'({e^c}_1)''+\frac{\theta^2}{\Lambda^{8}}(H_d)'(l\varphi_T)({e^c}_1)''
\label{eqn:model2.3}
\end{eqnarray}

\subsection{Charged Fermion Mass Matrices}
\label{sec:massmatrices}
The Higgs multiplets obtain their VEVs along with the $A_4$ and $\U(1)$ flavons $\varphi_T,\theta'',\theta$ 
as in Eqs.\ref{phi}-\ref{theta} 
leading to 
mass matrices of the following form:
\begin{equation}
 m_u\sim\begin{pmatrix}
      (s_1s_2^3t''t^3+s_1^4{t''}^4)s_2^2 & (s_1^3{t''}^3+s_2^3t^3)s_1s_2 & s_1^2{t''}^2s_2 \\
      (s_1^3{t''}^3+s_2^3t^3)s_1s_2 & s_1^2{t''}^2s_1^2 & s_1t''s_1 \\
      s_1^2{t''}^2s_2 & s_1t''s_1 & 1
     \end{pmatrix}s_1v_u,
\label{eqn:upmatrix}
\end{equation}

\begin{equation}
\label{eqn:downmatrix}
m_d\sim\begin{pmatrix}
      s_2^3t^2 & s^2_2s_1t''t & s_2s_1^2{t''}^2 \\
       \dots & s_1s_2t & s^2_1t'' \\
      \dots & \dots & 1
     \end{pmatrix}s_1s_2v_Tv_d,
\end{equation}

\begin{equation}
\label{eqn:lepmatrix}
m_e\sim\begin{pmatrix}
      s_2^3t^2 & \dots & \dots \\
      s_2^2s_1t''t & s_2s_1t & \dots \\
      s_2s_1^2{t''}^2 & s^2_1t'' & 1
     \end{pmatrix}s_1v_Tv_d,
\end{equation} 

The dots in $m_d$ and $m_e$ are from higher order corrections to the vev of the 
$\varphi_T$ flavon alignment. Such corrections come from the heavier modes which 
have a higher mass through orbifolding and will alter the alignment 
of $\varphi_T$ as discussed in sec. \ref{sec:alignment}.

We set $s_1=\lambda$ and $s_2=\lambda^{3/2}$ with $\lambda=0.22$, we choose for simplicity 
$t=t''=\mathcal{O}(1)$. We should make clear that taking $t=t''=\mathcal{O}(1)$ means 
that we are not using the Froggatt-Nielsen mechanism to provide the suppression. Insted 
the hierarchies originate from the bulk suppression factors $s_i$. The mass matrices are then given by:
\begin{equation}
 m_u\sim\begin{pmatrix}
      \lambda^{7} & \lambda^{5.5} & \lambda^{3.5} \\
      \lambda^{5.5} & \lambda^{4} & \lambda^{2} \\
      \lambda^{3.5} & \lambda^{2} & 1
     \end{pmatrix}\lambda v_u.
\end{equation}
The down sector matrix is given by,
\begin{equation}
\label{eqn:downmatrix3}
m_d\sim\begin{pmatrix}
      \lambda^{4.5} & \lambda^{4} & \lambda^{3.5} \\
      \dots & \lambda^{2.5} & \lambda^{2} \\
      \dots & \dots & 1
     \end{pmatrix}\lambda^{2.5} v_Tv_d,
\end{equation}
where again the dots represent contributions from the corrections to the 
vacuum alignment. 
The charged lepton mass matrix is given by 
\begin{equation}
 \label{eqn:lepmatrix3}
m_{charged\ lepton}\sim\begin{pmatrix}
      \lambda^{4.5} & \dots & \dots \\
      \lambda^{4} & \lambda^{2.5} & \dots \\
      \lambda^{3.5} & \lambda^{2} & 1
     \end{pmatrix}\lambda^{2.5} v_Tv_d,
\end{equation}
In this model since the first two families are doubled, because the gauge breaking 
orbifolding makes half of the $\mathbf{10}$-plets heavy the, GUT relation 
$m_{\mathrm{down}}=m_{\mathrm{charged\ lepton}}^T$ for the first two families is not valid.

These mass matrices give us approximate quark masses and mixing angles of the correct order of magnitude.
For example the quark mixing angles are given roughly by,
\begin{eqnarray}
 \theta_{12}=\mathcal{O}(\lambda^{1.5})\\
 \theta_{23}=\mathcal{O}(\lambda^{2})\\
 \theta_{13}=\mathcal{O}(\lambda^{3.5}).
\end{eqnarray}

So far we have not specified the size of $v_T$ and $v_S$, However from the 
ratio of the top and bottom quark masses we expect 
\begin{eqnarray}
 \frac{m_b}{m_t}&=&\lambda^{3/2}\frac{v_d}{v_u}v_T\sim\lambda^2\nonumber\\
 \Rightarrow v_T&\sim&\frac{\lambda^{1/2}}{\tan\beta}\sim \frac{1}{2\tan\beta}
\end{eqnarray}
where $\frac{v_d}{v_u}=\tan\beta$.


\subsection{Neutrino sector}
\label{sec:neutrino}

In the neutrino sector the right-handed neutrino $A_4$ 
triplets live at the fixed point. The $\varphi_S$ lives in the $z_1$ direction along with the $A_4$ singlet flavon $\xi$. 
After these flavons develop a vev the gauge singlets $N$ become heavy and the seesaw mechanism takes place similar to 
\cite{Altarelli:2008bg},\cite{Burrows:2009pi} with the alteration that a zero vev $A_4$ singlet flavon is no longer 
required as the vacuum alignment is determined by boundary conditions rather than by the use of driving fields.
Thus we have,

\begin{equation}
 w_\nu \sim\frac{y^D}{\Lambda}H_u(Nl)+\frac{1}{\Lambda}x_a\xi(NN)+\frac{x_b}{\Lambda}\varphi_S(NN)
\end{equation}

After the fields develop VEVs, the gauge
singlets N become heavy and the see-saw mechanism takes place as
discussed in detail in \cite{Chen:2009um}, leading to the effective mass matrix for the
light neutrinos:
\begin{equation}
 m_\nu\sim\frac{1}{3a(a+b)}\begin{pmatrix}
                         3a+b & b & b \\
                        b & \frac{2ab+b^2}{b-a} & \frac{b^2-ab-3a^2}{b-a}\\
                        b & \frac{b^2-ab-3a^2}{b-a} & \frac{2ab+b^2}{b-a}
                        \end{pmatrix}\frac{{s_1(v_u)^2}}{\Lambda}
\end{equation}
where
\begin{equation*}
 a\equiv\frac{2x_as_1u}{(y^D)^2},b\equiv \frac{2x_bs_1v_S}{(y^D)^2}.
\end{equation*}
The neutrino mass matrix is diagonalised by the transformation
\begin{equation*}
 U_{\nu}^Tm_\nu U_{\nu}=\mathrm{diag}(m_1,m_2,m_3)
\end{equation*}
with $U_{\nu}$ given by:
\begin{equation*}
 U_{\nu}=\begin{pmatrix}
    -\sqrt{2/3} & 1/\sqrt{3} & 0 \\
     1/\sqrt{6} & 1/\sqrt{3} & 1/\sqrt{2} \\
     1/\sqrt{6} & 1/\sqrt{3} & -1/\sqrt{2}
   \end{pmatrix}
\end{equation*}
which is of the TB form in Eq.~(\ref{MNS0}). However, although we
have TB neutrino mixing in this model we do not have exact TB
lepton mixing due to fact that the charged lepton mass matrix is
not diagonal in this basis. Thus there will be charged lepton
mixing corrections to TB mixing resulting in mixing sum rules as
discussed in \cite{King:2005bj,Antusch:2008yc}.

\subsection{Higher order corrections}

We will now discuss corrections to the mass matrices, such corrections 
come from additional flavon insertion of $\varphi_T,\varphi_S,\xi$ and $\theta,\theta''$, 
and also from corrections to the vacuum alignment of the $A_4$ triplet flavons $\varphi_T$ and $\varphi_S$.

\subsubsection{corrections to $m_{\mathrm{up}}$}

The leading order terms in the up sector are of the form
$\theta^m\theta''^nH_uq_iu_j$. Terms are gauge and $A_4$ singlets,
to create higher order terms we need to introduce flavon fields.
The most straightforward way to do this is to introduce two flavon
fields $(\varphi_T\varphi_T)_{\mathbf{1}}$, since $\varphi_T$ is an $A_4$ triplet
we need the two triplet fields in order to construct an $A_4$ singlet. Such terms
will lead to entries in the mass matrix suppressed by a factor of
$s_2^2v_T^2$. Due to the $\mathbb{Z}_3$ symmetry the flavon fields
$\varphi_S,\xi,\tilde{\xi}$ must enter at the three flavon level
so entries will be suppressed by a factor of $s_1^3v_S^2u,s_1^3v_S^3$ and
$s_1^3u^3$ relative to the leading order term. Using the values assumed in sec. \ref{sec:massmatrices} 
the corrections enter at $\mathcal{O}(\lambda^3)$ relative to the leading order term. 

\subsubsection{corrections to $m_d$ and $m_e$}
In the down quark mass matrix sub-leading corrections fill in the entries
indicated by dots in Eq. \ref{eqn:downmatrix}. Entries in the matrix are generated by terms of
the form $ \theta^m{\theta''}^nH'_d((d^c\varphi_T)q_i+(l\varphi_T)e^c_i)$,
higher order terms can come from replacing $\varphi_T$ with a
product of flavon fields or including the effect of the
corrections to the VEV of $\varphi_T$. 

The obvious substitution is to replace $\varphi_T$
with $\varphi_T\varphi_T$, this is compatible with the
$\mathbb{Z}_3$ charges and results in corrections with the same
form as $m_{\mathrm{down}}$ but with an extra overall suppression
of $s_2v_T$. Using the values assumed in sec. \ref{sec:massmatrices} this 
type of correction enters at the level of $\mathcal{O}(\lambda^{3/2})$ 

If we include the corrections to the alignment of the VEV of $\varphi_T$
then we fill in the entries indicated by dots in Eq. 
(\ref{eqn:downmatrix}). Such corrections originate from higher,heavy modes 
of the flavon field $\varphi_T$, such corrections would be suppressed by an 
order of $s_2$ relative to the leading order term giving corrections to 
the mass matrix of the form:
\begin{equation}
\delta m_{\mathrm{down}}\sim \begin{pmatrix}
      \lambda^{5} & \lambda^{5} & \lambda^{5} \\
      \lambda^{3.5} & \lambda^{3.5} & \lambda^{3.5} \\
      \lambda^{1.5} & \lambda^{1.5} & \lambda^{1.5}
     \end{pmatrix}\lambda^{2.5} v_Tv_d,  
\end{equation}
i.e. the corrections are suppressed by $\mathcal{O}(\lambda^{3/2})$ relative to the 
largest term in each row (or column for $m_{\mathrm{charged\ lepton}}$).

As remarked, since the first two families are doubled, because the gauge breaking 
orbifolding makes half of the $\mathbf{10}$-plets heavy the, GUT relation 
$m_{\mathrm{down}}=m_{\mathrm{charged\ lepton}}^T$ for the first two families is not valid. It 
does however hold up to orders of magnitude for the individual 
families so that the power of $\lambda$ is the same for each 
family though the (suppressed) $\mathcal{O}(1)$ coefficient can be 
different for each family.  

\subsubsection{corrections to $m_\nu$}
The leading order Dirac mass term for the neutrinos is $H_u(Nl)$, sub-leading corrections to this 
term enter with a single flavon insertion of $\varphi_T$ so the resulting term is $H_u(\varphi_TNl)$ 
this results in the sub-leading corrections entering at the $s_2v_T$ level.Using the values 
assumed in sec. \ref{sec:massmatrices} the corrections enter at the $\mathcal{O}(\lambda^{3/2})$ level. 

Corrections to the Majorana mass matrix can arise from a number of terms. This is 
due to the term $(NN)$ being a product of two triplets and can thus be a triplet 
or any of the singlet representations of $A_4$. Corrections to the Majorana mass matrix 
can have one extra flavon insertion relative to the leading order terms $\xi(NN),(\varphi_SNN)$. 
For example the term $(\varphi_S\varphi_T)(NN)$ is allowed by the $\mathbb{Z}_3$ symmetry and 
leads to corrections of order $s_2v_T$. After the seesaw mechanism 
takes place corrections to the neutrino masses and tri-bimaximal mixing are of order $s_2v_T$. 
Using the values assumed in sec. \ref{sec:massmatrices} these corrections are $\mathcal{O}(\lambda^{3/2})$ 
relative to the leading order term.


\section{Conclusion}
\label{sec:conclusion}

We have proposed the first realistic $\mathcal{N}=1$ SUSY $\SU(5)$ GUT model in 
8 dimensions with an $A_4$ family symmetry where the vacuum alignment is straightforwardly 
achieved by the use of boundary conditions on orbifolds of the four compact
dimensions. The low energy theory is the usual $\mathcal{N}=1$ SUSY Standard Model  
in 4 dimensions but with predictions for quark and lepton (including neutrino) masses and mixing angles. 
For example, the low energy 4d model naturally has TB mixing at the
first approximation and reproduces the correct mass hierarchies
for quarks and charged leptons and the CKM mixing pattern. The
presence of $SU(5)$ GUTs means that the charged lepton mixing
angles are non-zero resulting in predictions such as lepton
mixing sum rules.

We were motivated to consider an 8d theory by the desire to
achieve the $A_4$ flavon vacuum alignment in an elegant way using orbifold boundary conditions. 
Such boundary conditions result in the required 
alignment surviving at the zero mode level, and in relatively small corrections to the 
alignment resulting from heavy higher modes.  However 
the extra dimensional set up also provides familiar added benefits 
such as orbifold gauge and SUSY breaking with doublet-triplet 
splitting of the $\mathbf{5}$ and $\bar{\mathbf{5}}$ Higgs 
multiplets, making the coloured triplets heavy. 
Because the first two generations of $\mathbf{10}$-plets 
are doubled, both unwanted and desirable GUT relations are also avoided. The lack of 
such relations introduces more freedom into the theory.
The specific model in in table \ref{table:particles} and 
figure \ref{fig:model}  also includes a Froggatt-Nielsen $U(1)$ symmetry,
which, together with the bulk suppression factors, leads to the desired
inter-family hierarchies. 

Finally we comment on the possible relation between the 8d orbifold GUT-Family model
considered here and string theory. At first glance there is an
intriguing similarity between the model  
here and the F-theory GUT recently discussed \cite{Heckman:2008qa}.
In both cases the $SU(5)$ GUT gauge group lives in the full 8d space, and also the
matter and Higgs fields lie on matter curves in a 6d subspace, 
corresponding to two extra complex dimensions $z_{1,2}$,
with Yukawa couplings occurring at a 4d point \cite{Heckman:2008qa}. 
However any possible connection would be more subtle than this, since 
firstly one must uplift the 8d orbifold GUT-Family model here into full heterotic string  
theory, then one must identify duality relations between the hererotic string theory
and F-theory as discussed in \cite{Raby:2009dm}.
Nevertheless the 8d orbifold GUT-Family model here may provide a useful stepping-stone towards 
some future unified string theory (including gravity, albeit perhaps decoupled
in some limit) in which GUT breaking and the emergence of family
symmetry, spontaneously broken with a particular 
vacuum alignment, can be naturally explained as the result of the
compactification of extra dimensions.

  \section*{Acknowledgements}
  SFK acknowledges the support of a Royal Society Leverhulme Trust Senior Research Fellowship and the STFC Rolling Grant ST/G000557/1.

\end{document}